\documentclass{PoS}

\title{Systematic study of 1-loop correction on sparticle decay widths using GRACE/SUSY-loop}

\ShortTitle{1-loop correction on sparticle decays with GRACE/SUSY-loop}

\author{\speaker{K.~Iizuka},$^a$  T.~Kon,$^a$ K.~Kato,$^b$ T.~Ishikawa,$^c$ Y.~Kurihara,$^c$ M.~Jimbo,$^d$ and M.~Kuroda $^e$\\
        \llap{$^a$} Seikei University, Musashino, Tokyo 180-8633, Japan\\
        \llap{$^b$} Kogakuin University, Shinjuku, Tokyo 163-8677, Japan \thanks{KU-PH-004}\\
        \llap{$^c$} KEK, Tsukuba, Ibaraki 305-0801, Japan \thanks{KEK-CP-228} \\
        \llap{$^d$} Chiba University of Commerce, Ichikawa, Chiba 272-8512, Japan \thanks{CUCP-10-1} \\
        \llap{$^e$} Meiji Gakuin University, Yokohama, Kanagawa 244-8539, Japan\\
        E-mail: \email{dm083502@cc.seikei.ac.jp}, \email{kon@st.seikei.ac.jp}}

\abstract{The 1-loop corrected decay widths of sparticles (charginos, neutralinos, gluino and sfermions) in the framework of the MSSM are calculated systematically using {\tt GRACE/SUSY-loop}, which is the program package for the automatic calculation of the MSSM amplitudes in the 1-loop order. We present the renormalization scheme used in our system and show some numerical results of decay widths of sfermions and gluino using the SPS1a' parameter set and other SUSY parameter sets. \
}

\FullConference{RADCOR 2009 - 9th International Symposium on Radiative Corrections (Applications of Quantum Field Theory to Phenomenology) ,\\
		 October 25 - 30 2009\\
		 Ascona, Switzerland}

\begin{document}

{\flushleft{\large 1. Introduction}}
\vskip5pt
The supersymmetric (SUSY) theory is a good candidate for the theory beyond the standard model. 
So the experimental confirmation of the SUSY theory is one of the most important themes of  the present and future particle experiments. 
Actually, discoveries of the SUSY particles (sparticles) are expected at LHC and ILC. 
Since we expect very accurate experimental data at ILC, we need theoretical prediction which match the measurement accuracy. 
Apparently, the tree-level calculation is insufficient.   
So we are calculating the radiative correction to possible major decay modes of sparticles 
using {\tt GRACE/SUSY-loop} \cite{ourpr}.
In this paper we report numerical results on the decays channels for squarks and gluino \cite{ghs,bhpz}, 
which are SUSY partners of quarks and gluon, respectively.
\vskip10pt
{\flushleft{\large 2. GRACE/SUSY-loop}}
\vskip5pt
Analytical evaluation of physical amplitudes characterized by many Feynman diagrams is not easy. 
It is principally for this reason that we have developed the {\tt GRACE} system \cite{gracehp}.
This system calculates the cross sections and the decay widths and generates events automatically 
in the following way, 
$(1)$ it generates all Feynman diagrams automatically, 
$(2)$ it generates physical amplitudes automatically,
$(3)$ it incorporates libraries (loop integral, kinematics, etc.), 
$(4)$ it integrates the matrix element by the adaptive Monte Carlo method, 
$(5)$ it generates Monte Carlo events, 
$(6)$ it has various self-test mechanisms of the results (UV and IR cancellation, NLG invariance etc.).
{\tt GRACE/SUSY-loop} can calculate the SUSY amplitudes up to 1-loop order. 
For the calculation of the SUSY  amplitudes at 1-loop level, there are also other programs, 
{\tt SloopS} \cite{sloops} and {\tt FeynArt/Calc} \cite{feynac}. 

In {\tt GRACE/SUSY-loop}, we have used the technique of the non-linear gauge (NLG) \cite{nlg} in order to test the system.
Concretely, we introduce the following gauge fixing terms in the Lagrangian. 
\begin{eqnarray}
&&F_{W^{\pm}}=(\partial_\mu \pm i e {\tilde{\alpha}} A_\mu 
  \pm i g \cos\theta_W {\tilde{\beta}} Z_\mu ) W^{\pm \mu} 
 \pm i \xi_W {\frac{g}{2}}(v + \tilde{\delta}_H H^0 
   + \tilde{\delta}_h h^0 \pm i \tilde{\kappa} G^0 ) G^\pm, \\
&&F_Z=\partial_\mu Z^\mu +
 \xi_Z {\frac{g_Z}{2}}(v + \tilde{\epsilon}_H H^0 
   + \tilde{\epsilon}_h h^0) G^0, \\
&&F_\gamma = \partial_\mu A^\mu.
\end{eqnarray}
They contain seven independent NLG-parameters, 
$({\tilde{\alpha}}, {\tilde{\beta}}, \tilde{\delta}_H, \tilde{\delta}_h, \tilde{\kappa}, \tilde{\epsilon}_H, \tilde{\epsilon}_h)$. 
We emphasize that the NLG interactions are included in the electroweak (ELWK) sector of the MSSM Lagrangian. 
While each Feynman diagram depends on the NLG-parameters, the sum of all diagrams should be independent of them. 
We can confirm the validity of calculation when the physical quantities do not change the value 
for different sets of numerical values of the NLG-parameters. 
It is the test of the gauge invariance.

We use the on-mass-shell conditions as much as possible for the renormalization of the ELWK sector. 
As a result, gauge bosons, all fermions $f$, sfermions $\widetilde{f}$ and 
the lightest neutralino $\widetilde{\chi}^0_1$ have 
no mass shifts in the ELWK 1-loop order. 
We should note that there are some freedom in the renormalization scheme of the sfermion sector. 
They are distinguished by different choice of residue conditions, decoupling conditions on the transition terms between lighter and heavier sfermions, and the left-handed SU(2) relations in the 1-loop order. 
In this paper the calculation is done with the scheme in which we impose the residue conditions on all sfermions except for heavier stop and sbottom ($\widetilde{t}_2$, $\widetilde{b}_2$). 
The external line corrections for these particles become non-zero in this scheme. 

Renormalization in the QCD sector is done in a mixed scheme. 
Light (1st and 2nd generation) quarks and gluon are treated in the ${\overline{DR}}$ scheme as in the convensional perturbative QCD. 
Massive particles are handled by the on-mass-shell scheme as in the ELWK sector. 
For the regularization of infrared divergences, the previous version \cite{ourpr} of the {\tt GRACE/SUSY} system used the fictitious mass of gluon $\lambda$. 
We have developed a new system in which mass-singularities are regularized by the dimensional method.
In order to refer the ultraviolet and infrared divergences we define the notations of $C_{UV}\equiv {1}/{\epsilon}$ and $C_{IR}\equiv {1}/{\bar{\epsilon}}$, where the dimension of the space-time $d = 4-2\epsilon = 4+2{\bar \epsilon}$. 
In the following numerical calculation, we mainly use the SPS1a' parameter set \cite{spa1}. 
\vskip10pt
{\flushleft{\large 3. Squarks and gluino decays}}
\vskip5pt
Possible decay modes of sfermions and gluino are as follows. 
\begin{eqnarray}
&&\widetilde{q} \to q \widetilde{\chi}^0_i, \qquad \widetilde{q} \to q' \widetilde{\chi}^+_k \qquad (i=1 \sim 4,\quad k=1, 2) \\
&&\widetilde{\ell} \to \ell \widetilde{\chi}^0_i, \qquad \widetilde{\ell} \to \ell' \widetilde{\chi}^+_k 
\qquad (i=1 \sim 4,\quad k=1, 2) , \\
&& \widetilde{g} \to q \widetilde{q}_j. \qquad  \qquad  \qquad \qquad  (j=1, 2)
\label{sgdc}
\end{eqnarray}
Note that squarks cannot decay into $q+\widetilde{g}$ because the gluino mass is larger than all squark masses in the SPS1a'. 
Gluino decays into most of quark$-$squark pairs, but it cannot decay into the top and the heavier stop 
$\widetilde{t}_2$ because of $m_{\widetilde{g}} < m_t + m_{\widetilde{t}_2}$. 

Among various squark decay channels here we focus on the lighter stop $\widetilde{t}_1$ decays.
From the tree level calculation, we find 
$Br(\widetilde{t}_1 \to b \widetilde{\chi}^+_1) = 86.7\%$ and 
$Br(\widetilde{t}_1 \to t \widetilde{\chi}^0_1) = 13.3\%$. 
In the following we show detailed result on the main mode $\widetilde{t}_1 \to b \widetilde{\chi}^+_1$. 
Twelve Feynman diagrams for the 1-loop electroweak correction among 82 diagrams are shown in Figure 1.

\begin{figure}[htbp]
\begin{center}
\includegraphics[width=.5\textwidth, angle=0]{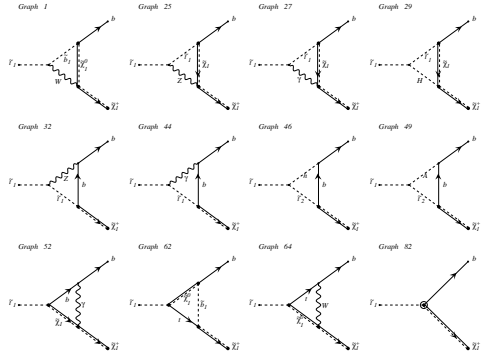}
\end{center}
\vspace{-1cm}
\caption{Twelve 1-loop electroweak Feynman diagrams for $\widetilde{t}_1 \to b \widetilde{\chi}^+_1$ 
among 82 diagrams. The 82th diagram is counter term.}
\label{fig1}
\end{figure}

Since {\tt GRACE} is an automatic calculation system, it is important to test reliability of the results. 
The numerical values obtained by the system must pass all following tests.
First, we show the independence of the NLG-parameters. If we change 7 gauge parameters, 
each evaluated value of loop graphs changes but the total  does not change. 
We compare the case 1 :  
$({\tilde{\alpha}}, {\tilde{\beta}}, \tilde{\delta}_H, \tilde{\delta}_h, \tilde{\kappa}, \tilde{\epsilon}_H, \tilde{\epsilon}_h)=
(0,0,0,0,0,0,0)$ with 
the case 2 : $(1000, 2000, 3000, 4000, 5000, 6000, 7000)$. 
We have obtained the sum of the loop contribution, $\delta\Gamma_{loop}$
 and soft-photon contribution, $\delta\Gamma_{soft}$ as follows (units in GeV).
\begin{eqnarray*}
{\rm case 1 : }&\quad&\delta\Gamma_{loop}+\delta\Gamma_{soft}=0.15117115752797127186610833503954323 \\
{\rm case 2 : }&\quad&\delta\Gamma_{loop}+\delta\Gamma_{soft}=0.15117115752797127186610833480863836
\end{eqnarray*}
Both numbers agree up to 26 digits. So, the gauge invariance is confirmed. 
Next, we show cancellation tests of ultraviolet (UV)  and Infrared (IR) divergence. 
We keep the UV divergent factor $C_{UV}$ and a tiny fictitious mass $\lambda$ of photon to regularize the IR divergence in the program. 
In Table 1,  we find that $\delta\Gamma_{loop}$ is the same for $C_{UV}=0$ and  $C_{UV}=1000$. 
As for the cancellation test of infrared divergence, we see the sum of $\delta\Gamma_{loop}$ and $\delta\Gamma_{soft}$ remain unchanged when we change $\lambda = 10^{-24}$ to $10^{-27}$(GeV).  
Finally, we can test that the sum of $\delta\Gamma_{loop}$, $\delta\Gamma_{soft}$ and 
the hard-photon contribution, $\delta\Gamma_{hard}$ is 
independent of the soft photon cut parameter $k_c$. 
Since the numerical values of the sum 
$\delta\Gamma_{ELWK}$ ($=$ $\delta\Gamma_{loop}$ $+$ $\delta\Gamma_{soft}$ $+$ $\delta\Gamma_{hard}$) 
are the same within accuracy,  the reliable value, 13.9\% correction is obtained.
\begin{table*}[htb]
\begin{small}
\begin{center}
\begin{tabular}{@{}cccccc}
\hline
&$C_{UV}$ & $0$ & $1000$ & $0$ & $0$ \\
&$\lambda$ (GeV)& $10^{-24}$ & $10^{-24}$ & $10^{-27}$ & $10^{-24}$ \\
&$k_c$ (GeV)& $10^{-3}$ & $10^{-3}$ & $10^{-3}$ & $10^{-5}$ \\
\hline
 & $\delta\Gamma_{loop}$ (GeV) &
   $-0.06256$ & $-0.06256$ & $-0.09364$ & $-0.06256$ \\
 & $\delta\Gamma_{soft}$ (GeV) & 
   $0.21373$ & $0.21373$ & $0.24481$ & $0.19301$ \\
 & $\delta\Gamma_{hard}$ (GeV) &
   $0.04849$ & $0.04849$ & $0.04849$ & $0.06921$ \\
 & $\delta\Gamma_{ELWK}$  (GeV) &
   $0.19966$ & $0.19966$ & $0.19966$ & $0.19966$ \\
 & $\delta\Gamma_{ELWK}/\Gamma_{tree}$  &
   $13.9\%$ & $13.9\%$ & $13.9\%$ & $13.9\%$ \\
\hline
\end{tabular}\\[2pt]
\end{center}
\end{small}
\vspace{-10pt}
\caption{ELWK 1-loop corrections to the $\widetilde{t}_1 \to b \widetilde{\chi}^+_1$ decay width for SPS1a'}
\label{table1}
\end{table*}

For the QCD correction (see Table 2), we use the ${\overline{DR}}$ scheme. 
Like the ultraviolet divergence factor $C_{UV}$, 
we keep the infrared divergent factor $C_{IR}$ in the calculation. 
Comparing various case of $(C_{UV},C_{IR}, k_c)$, we obtain consistent and reliable value, $-$7.1\% correction.
In the {\tt GRACE} system, we can also calculate the correction with the fictitious gluon mass 
and obtain the same correction $-$7.1\%.

\begin{table*}[htb]
\begin{small}
\begin{center}
\begin{tabular}{@{}cccccc}
\hline
&$C_{UV}$ & $0$ & $1$ & $0$ & $0$ \\
&$C_{IR}$ & $0$ & $0$ & $1$ & $0$ \\
&$k_c$(GeV) & $10^{-3}$ & $10^{-3}$ & $10^{-3}$ & $10^{-4}$ \\
\hline
 & $\delta\Gamma_{loop}$ (GeV)  &
   $-1.254$ & $-1.254$ & $-1.479$ & $-1.254$ \\
 & $\delta\Gamma_{soft}$ (GeV)  & 
   $-3.752$ & $-3.752$ & $-3.527$ & $-4.786$ \\
 & $\delta\Gamma_{hard}$ (GeV)  &
   $4.905$ & $4.905$ & $4.905$ & $5.939$ \\
 & $\delta\Gamma_{QCD}$  (GeV)   &
   $-0.100$ & $-0.100$ & $-0.100$ & $-0.099$ \\
 & $\delta\Gamma_{QCD}/\Gamma_{tree}$  &
   $-7.1\%$ & $-7.1\%$ & $-7.1\%$ & $-7.1\%$ \\
\hline
\end{tabular}\\[2pt]
\end{center}
\end{small}
\vspace{-10pt}
\caption{QCD 1-loop corrections to the $\widetilde{t}_1 \to b \widetilde{\chi}^+_1$ decay width for SPS1a'}
\label{table2}
\end{table*}

Adding the ELWK and QCD corrections, we obtain 
$\delta\Gamma\left(\widetilde{t}_1 \to b \widetilde{\chi}^+_1 \right)/\Gamma_{tree}$ $=$ $13.9\%$ $-7.1\%$ $=$ $6.8\%$, 
where $\Gamma_{tree} = 1.43$GeV.
Similarly we obtain the results for the other channel, 
$\delta\Gamma\left(\widetilde{t}_1 \to t \widetilde{\chi}^0_1 \right)$ $/\Gamma_{tree}$ $=$ $7.6\%$ $+2.1\%$ $=$ $9.7\%$, 
where $\Gamma_{tree} = 0.22$GeV.
Fig.~2 shows the branching ratio at the tree and the 1-loop level. 
There is only a small difference between them.  

\begin{figure}[htbp]
 \begin{minipage}{0.5\hsize}
  \begin{center}
   \includegraphics[width=.6\textwidth, angle=90]{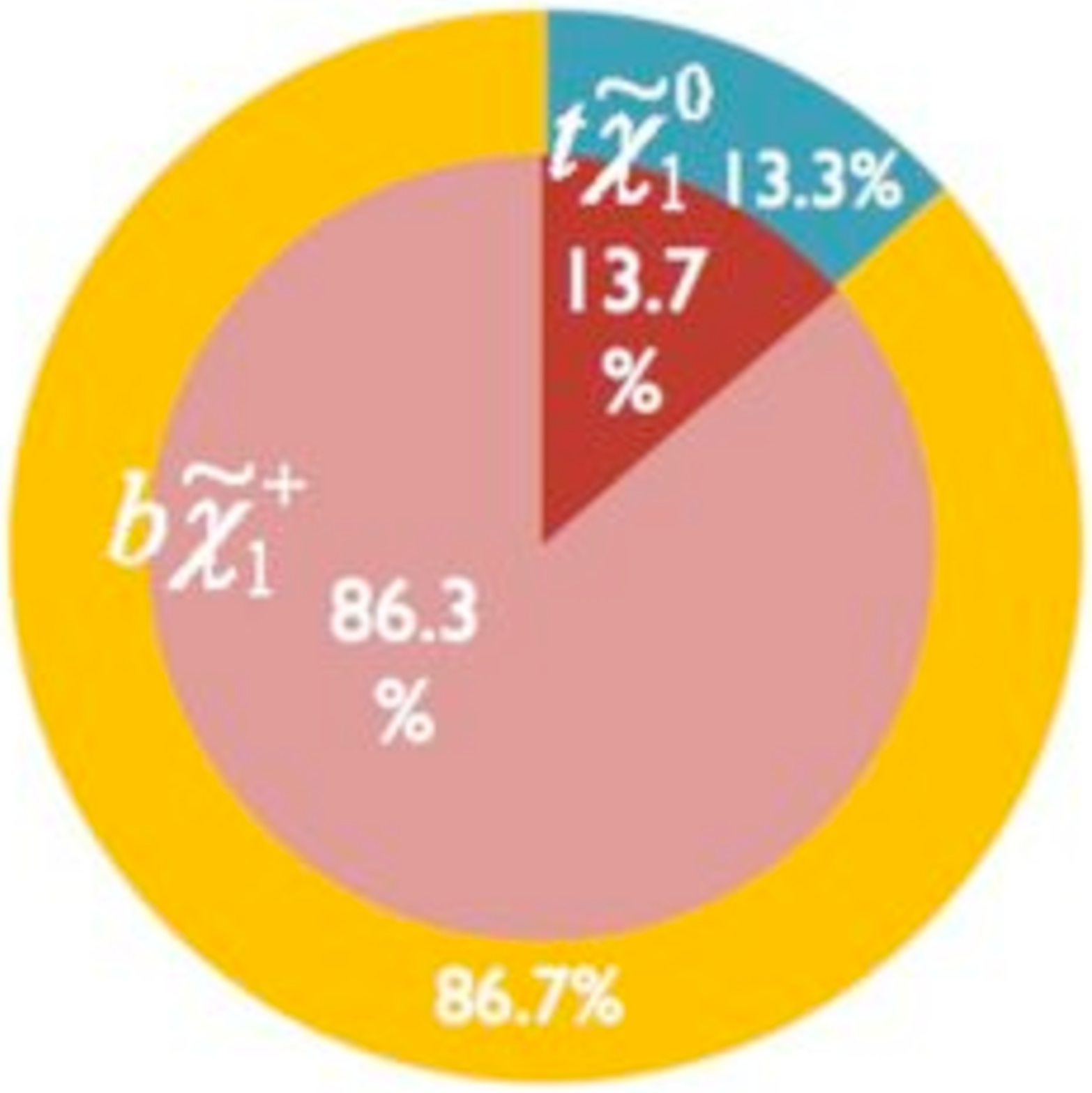}
\end{center}
\vspace{-10pt}
\caption{Branching ratios of $\widetilde{t}_1$ decay. Outer circle and inner one correspond to tree level and 1-loop corrected, respectively.}
\label{fig2}
 \end{minipage}
\begin{minipage}{0.5\hsize}
\begin{center}
\vspace{-15pt}
\includegraphics[width=.8\textwidth]{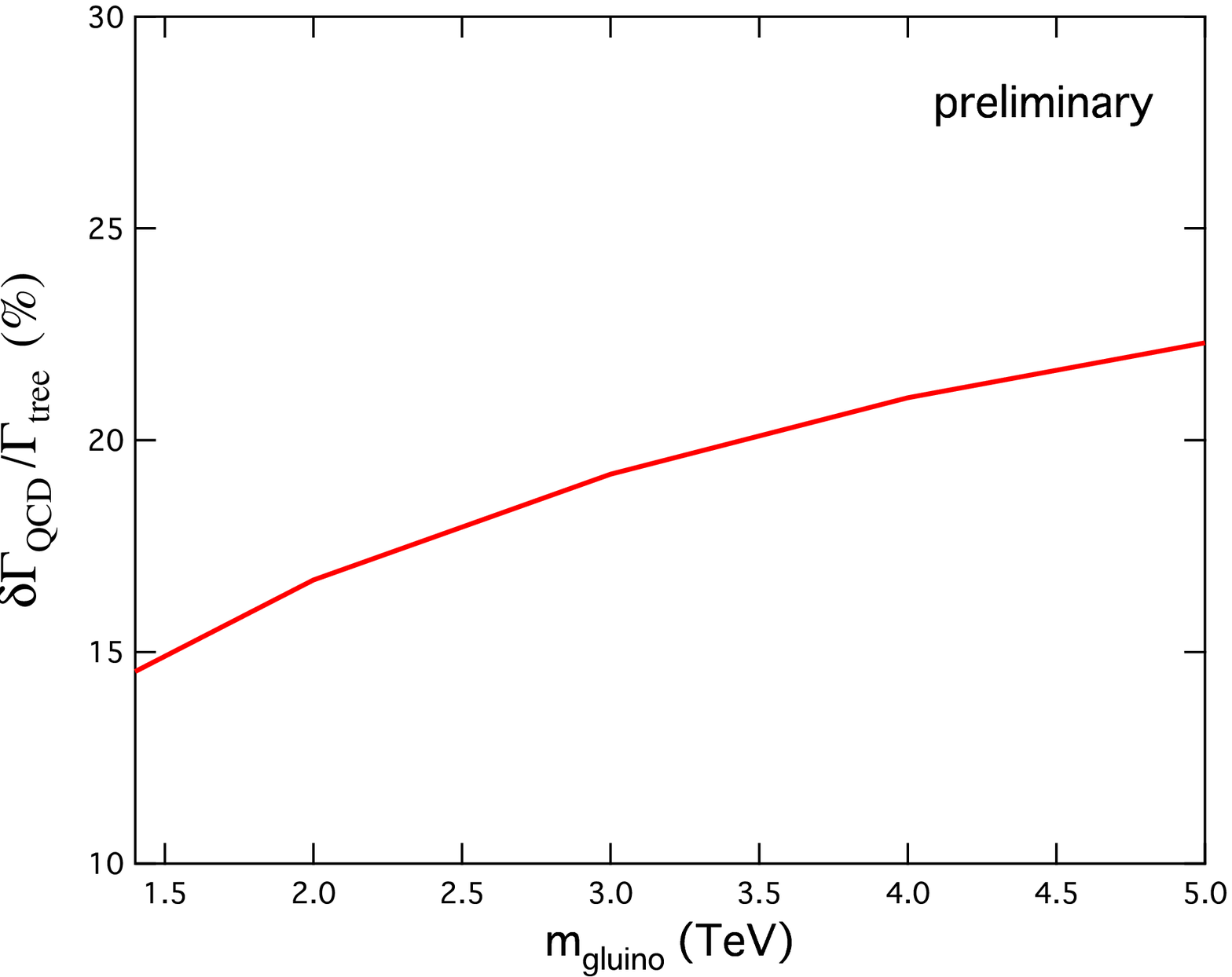}
\end{center}
\vspace{-0.5cm}
\caption{$m_{\widetilde{g}}$ dependence of ${\frac{\delta\Gamma_{QCD}}{\Gamma_{tree}}}(\widetilde{t}_1 \to b W^+ \widetilde{\chi}^0_1)$}
\label{fig3}
 \end{minipage}
\end{figure}

Up to now, we have considered the SPS1a' parameter set. Now we take another MSSM parameter set in which 
two-body decay channels $\widetilde{t}_1 \to t \widetilde{\chi}^0_1$ and $\widetilde{t}_1 \to b \widetilde{\chi}^+_1$
of the lighter stop $\widetilde{t}_1$ is kinematically forbidden and $\widetilde{t}_1$ 
dominantly decays into 3-body channel, 
$\widetilde{t}_1 \to b W^+ \widetilde{\chi}^0_1$. 
This is the case when the following mass relations hold,   
$m_{\widetilde{t}_1} > m_b + m_W + m_{\widetilde{\chi}^0_1}$, 
$m_{\widetilde{t}_1} < m_t + m_{\widetilde{\chi}^0_1}$ and 
$m_{\widetilde{t}_1} < m_b + m_{\widetilde{\chi}^+_1}$. 
As an example, here we take 
$m_{\widetilde{t}_1} = 300$GeV, $m_{\widetilde{\chi}^0_1} = 195$GeV, $m_{\widetilde{\chi}^+_1} = 396$GeV and obtain $\Gamma_{tree} = 0.664$keV. 
Through the gluino$-$squark loop contributions,  
the corrected width depends on the gluino mass as well as masses of the 1st and 2nd generation squarks. 
In Fig.~3 the gluino mass dependence of the correction ${\frac{\delta\Gamma_{QCD}}{\Gamma_{tree}}}$ are shown. 
We find that 2TeV gluino mass shift induces about 5\% shift of the width.
If the gluino $\widetilde{g}$ is too heavy to be produced at future colliders, the precision measurements of the decay width of 
the light stop $\widetilde{t}_1$ will give us information of the gluino. 


Similarly, we have calculated the 1-loop ELWK and QCD correction of gluino decays (\ref{sgdc}). 
As for the QCD correction, our results agree with the calculation done by Beenakker et al. \cite{bhpz}, 
if we use the same input values as theirs (see Fig.~4). 
\begin{figure}[htbp]
\begin{center}
\includegraphics[width=.4\textwidth, angle=0]{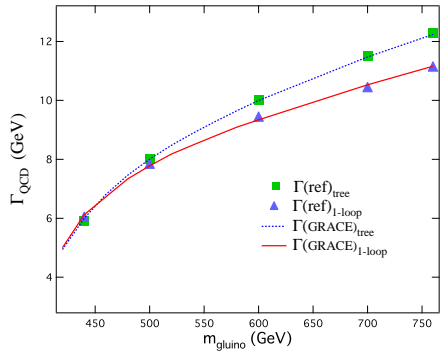}
\end{center}
\vspace{-20pt}
\caption{Gluino mass dependence of QCD correction to $\widetilde{g} \to t \widetilde{t}_1$. Dotted line and solid line respectively 
corresponds to $\Gamma_{\rm tree}$ and $\Gamma_{\rm 1-loop}$ calculated by {\tt GRACE}. Rectangle and triangle corresponds to 
$\Gamma_{\rm tree}$ and $\Gamma_{\rm 1-loop}$ shown in Fig.~5 of the reference \cite{bhpz}.}
\label{fig4}
\end{figure}
When we take the SPS1a' parameter set, we obtain 
\begin{eqnarray}
&&\delta\Gamma/\Gamma_{tree}(\widetilde{g} \to b \widetilde{b}_1)=-18.0\%({\rm QCD})+2.5\%({\rm ELWK}) \\
&&\delta\Gamma/\Gamma_{tree}(\widetilde{g} \to t \widetilde{t}_1)=-13.0\%({\rm QCD})+1.2\%({\rm ELWK}).
\end{eqnarray}

\vskip10pt
{\flushleft{\large 4. Summary}}
\vskip5pt
Since the sparticles are expected to have masses of the order of the ELWK scale, we cannot neglect the ELWK corrections 
as well as the QCD corrections in the precise theoretical prediction of their production cross sections and the decay rates. 
Using {\tt GRACE/SUSY-loop}, we have systematically calculated ELWK and QCD corrections to the sfermion and gluino decays, 
and confirmed that both corrections are equally important. 
We have found that QCD corrections for $\widetilde{t}_1 \to b \widetilde{\chi}^+_1$ is the same 
in the ${\overline{DR}}$ regularization and in the regularization with the fictitious gluon mass. 
We have already calculated the chargino \cite{ourpr} and the neutralino decay channels and have got reliable results. 
Extension of the adaptive range to radiative corrections for multi-body channels is now planned. 

\vspace{10pt}
This work is partially supported by Grant-in-Aid for Scientific Research(B)
(20340063) and Grant-in-Aid for Scientific Research on Innovative Areas (21105513).


\begin{thebibliography}{99}
\bibitem{ourpr} 
J.~Fujimoto, T.~Ishikawa, M.~Jimbo, T.~Kon, Y.~Kurihara and M.~Kuroda, 
\emph{Phys. Rev.} {\bf D75} (2007) 113002, [{\tt hep-ph/0701200}].
\bibitem{ghs} 
J.~Guasch, W.~Hollik and J~.Sola, 
\emph{JHEP} {\bf 0210} (2002) 040, [{\tt hep-ph/0207364}] ; 
\emph{Phys.Lett.} {\bf B437} (1998) 88, [{\tt hep-ph/9802329}] ;
\emph{"Electroweak radiative corrections to sfermion decays"}, LC-TH-2003-033, [{\tt hep-ph/0307011}]
\bibitem{bhpz} 
W.~Beenakker, R.~Hopker and P.~M.~Zerwas, 
\emph{Phys. Lett.} {\bf B378} (1996) 159,  [{\tt hep-ph/9602378}] ; \\
W.~Beenakker, R.~Hopker, T.~Plehn and P.~M.~Zerwas, 
\emph{Z. Phys.} {\bf C75} (1997) 349,  [{\tt hep-ph/9610313v1}].
\bibitem {gracehp} 
F.~Yuasa {\it et al.},
\emph{Prog. Theor. Phys. Suppl.} {\bf 138} (2000) 18, [{\tt hep-ph/0007053}] ; \\
{\tt http://minami-home.kek.jp/}
\bibitem{sloops}
N.~Baro and F.~Boudjema, 
\emph{Phys.Rev.} {\bf D80} (2009) 076010, [{\tt hep-ph/0906.1665}].
\bibitem{feynac}
T. Hahn, \emph{ Nucl. Phys. Proc. Suppl.} {\bf 89} (2000), 231 ;
\emph{ Comput. Phys. Commun.} {\bf 140} (2001), 418.
\bibitem{nlg} 
J.~Fujimoto, T.~Ishikawa, M.~Jimbo, T.~Kaneko, T.~Kon, Y.~Kurihara, M.~Kuroda and Y.~Shimizu, 
\emph{Nucl. Phys.} (Proc. Suppl.) 157 (2006) 157 ; \\
G.~B\'elanger, F.~Boudjema, J.~Fujimoto, T.~Ishikawa, T.~Kaneko, K.~Kato, Y.~Shimizu, 
\emph{Phys. Rep.} {\bf 430} (2006) 117, [{\tt hep-ph/0308080}].
\bibitem {spa1}  
J.A.~Aguilar-Saavedra {\it et al.}, \emph{Eur. Phys. J.} {\bf C46} (2006) 43.

\end{thebibliography}
\end{document}